\documentclass[12pt]{iopart}
\usepackage{graphicx}
\begin{document}

\title{Jet Correlations of Identified Particles in PHENIX}

\author{Anne Sickles\dag\ for the PHENIX Collaboration
\footnote[3]{For the full PHENIX Collaboration author list
and acknowledgments, see Appendix ``Collaborations'' of this volume.}
}

\address{\dag\ Department of Physics and Astronomy,
State University of New York at Stony Brook, Stony Brook NY 11794-3800, USA}

\begin{abstract}
Azimuthal two particle correlations at intermediate $p_{T}$ 
with one of the particles identified have been measured at
PHENIX.  Trigger ($2.5 < p_{T} < 4.0 GeV/c$) baryons
and mesons show 
little significant difference in the number
of associated particles ($1.7 < p_T < 2.5GeV/c$) independent
of centrality.  For inclusive hadron triggers
with $2.5 < p_{T} < 4.0GeV/c$, associated fragmentation
particles with $1.0 < p_T < 2.5GeV/c$ show a higher
baryon to meson ratio on the away side.
\end{abstract}




In Au+Au collisions at RHIC an anomalously large number of
$p$ and $\bar p$ have been observed at intermediate
$p_{T}$ ($2-5GeV/c$) in central collisions
while the $p/\pi$ ratio in peripheral collisions
is similar to p+p \cite{phpro}.  Coalescence/recombination models
\cite{fries,hwa,ko}
explain this ratio by modeling the production of baryons and
mesons in this $p_{T}$ range as coming dominantly from 
the coalescence of flowing thermal partons from
lower $p_{T}$ rather than the fragmentation of a
higher $p_{T}$ parton, as is usually assumed. 
Such models are able to reproduce the observed hadron
spectra and the species dependence of the elliptic
flow.
Angular correlations provide an important means
for testing this assumption.  If the intermediate $p_T$
hadrons are produced by fragmentation they should have
associated lower $p_T$ particles from the jet
fragmentation in a manner similar
to what is observed in p+p collisions.  If they, instead,
are produced by coalescence of thermal partons, there should be no
jet-like associated lower $p_T$ particles.
  We identify the trigger
particle as a baryon ($p$ or $\bar{p}$) or a meson 
($\pi$ or $K$) to investigate whether the 
excess baryons have a jet-like nature.

The particle composition of the fragmentation products,
i.\ e.\ lower $p_{T}$ particles associated with jets,
provides information about possible modification
of the fragmentation process by the hot matter created
in central Au+Au collisions.  By triggering on a leading
hadron, a bias is created toward the hard scattering occurring
such that the trigger particle traverses very little
of the hot matter.  Consequently the away side
jet at $\Delta\phi = \pi$ has a longer than average path length 
through the medium.  Differences in the particle composition 
of the associated particles on 
the away side compared to the near side of the jet provide 
information about the possible modification of the fragmentation
function in the hot medium.
\begin{figure}
 \begin{center}
   \includegraphics[width=8.0cm]{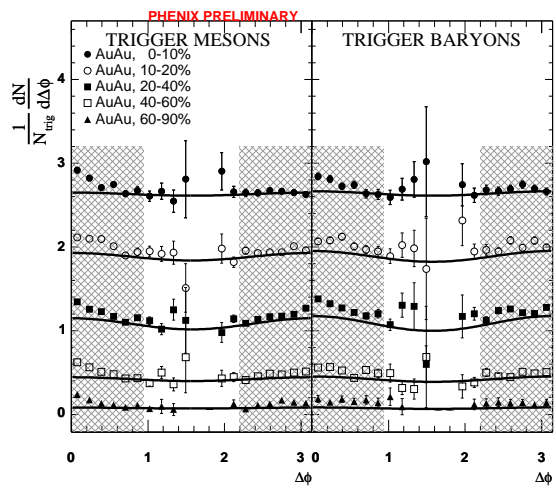}
   \caption{Fully corrected relative angle distributions for identified
	trigger particles for five centrality bins.}
   \label{idtrigdndp}
  \end{center}
\end{figure}

In this analysis 25 million minimum bias events are used.
Charged particles are tracked in both PHENIX arms by a 
Drift Chamber (DC) and Pad Chamber 1 (PC1). 
Particles are either  {\sl trigger particles} with
$2.5 < p_T < 4.0GeV/c$, a matching cut in the PHENIX
High Resolution Time of Flight (TOF), and a mass cut identifying
it as either a baryon or a meson ($\pi$ or $K$) or 
{\sl associated particles} with $1.7 < p_T < 2.5GeV/c$
and a matching cut in the Pad Chamber 3 (PC3).  
Particle correlations are studied as a function of $\Delta \phi$, the 
azimuthal angle difference between trigger and associated particles. 
The same distribution is made taking  trigger and associated particles 
from different events; the shape of the mixed event distribution provides 
a measure of the PHENIX azimuthal acceptance for particle pairs, and the 
absolutely normalized mixed events determine the combinatorial background. 
The associated particle yield per trigger is determined, correcting for 
the single track reconstruction efficiency. The resulting 
$1/N_{trig}dN/d\Delta\phi$ distribution 
has two sources of azimuthal correlations, elliptic flow and jets.
Elliptic flow produces an angular correlation of: 
\begin{equation}
$$B(1+2v_2^{assoc}v_2^{trig}\cos(2\Delta\phi))$$
\end{equation}
 where $v_2^{assoc}$ and $v_2^{trig}$ are the $v_2$ values
for the associated and trigger particles, respectively, 
taken from \cite{phenixv2}.    
$B$ is the level of combinatoric background and is
taken from the mixed events.  The combinatorial
background, with the elliptic flow
modulation is then subtracted, leaving only jet-like
correlations.  
Figure \ref{idtrigdndp} shows the $1/N_{trig}dN/d\Delta\phi$ 
distributions for five centralities with the solid 
lines indicating the non-jet correlations.
To get the near side (far side)
associated particle yield per trigger
we integrate the jet correlation $0 < \Delta\phi < 0.94$ 
($2.2 < \Delta\phi < \pi$).

In the second analysis, both particles are charged particles which are
tracked in the DC and PC1.
Trigger particles have $2.5 < p_T < 4.0GeV/c$, while
associated particles have $1.0 < p_T < 2.5GeV/c$ and
are identified as either protons/anti-protons or mesons
($\pi$, $K$) via timing in the PHENIX Electromagnetic Calorimeter (EMCal).
  The pairs are used to form traditional correlation 
functions, with the various contributions
deconvoluted via a simultaneous fit.  The $v_2$ values are obtained by fitting
correlation functions made with different orientations
of the trigger particle with respect to the reaction plane \cite{v2react}.
With the $v_2$ values known it is possible to fit the 
non-flow part of the correlation function to two Gaussians,
one for the near side jet and one for the away side jet.
Corrections for the partner efficiency are applied and the
absolute associated particle yield per trigger
for the near (away) side is the integral
of the Gaussian for $0 < \Delta\phi < \pi/2$ 
($\pi/2 < \Delta\phi < \pi$).

\begin{figure}
   \begin{minipage}{8.0cm}
   \includegraphics[width=8.0cm]{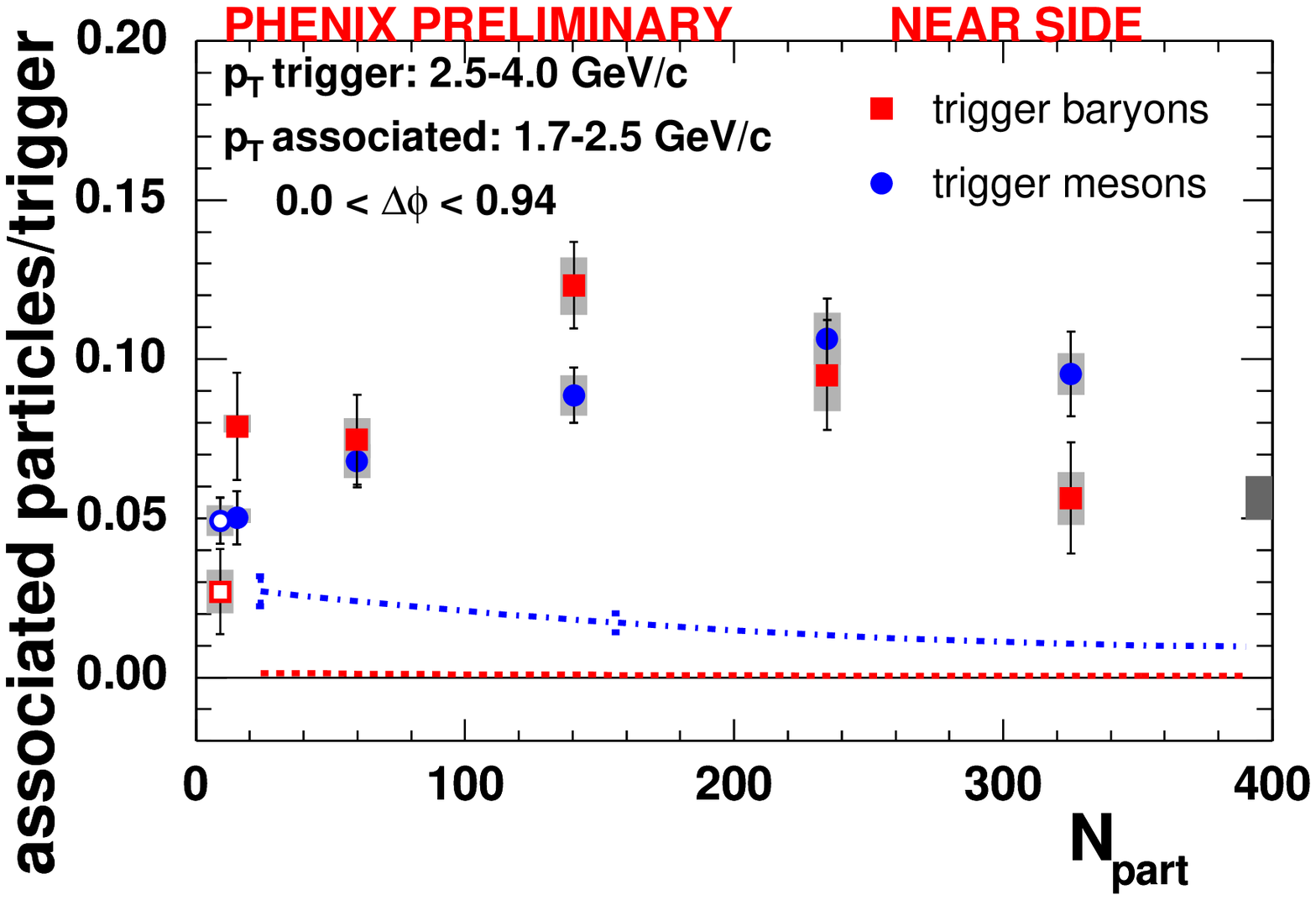}
   \caption{Near side associated particle yield per trigger. Dashed (dot-dashed) line is the prediction for protons (pions) from \cite{fries} as described in text.  Open points are d+Au values.}
   \label{idtrign}
   \end{minipage}
   \begin{minipage}{8.0cm}
   \includegraphics[width=8.0cm]{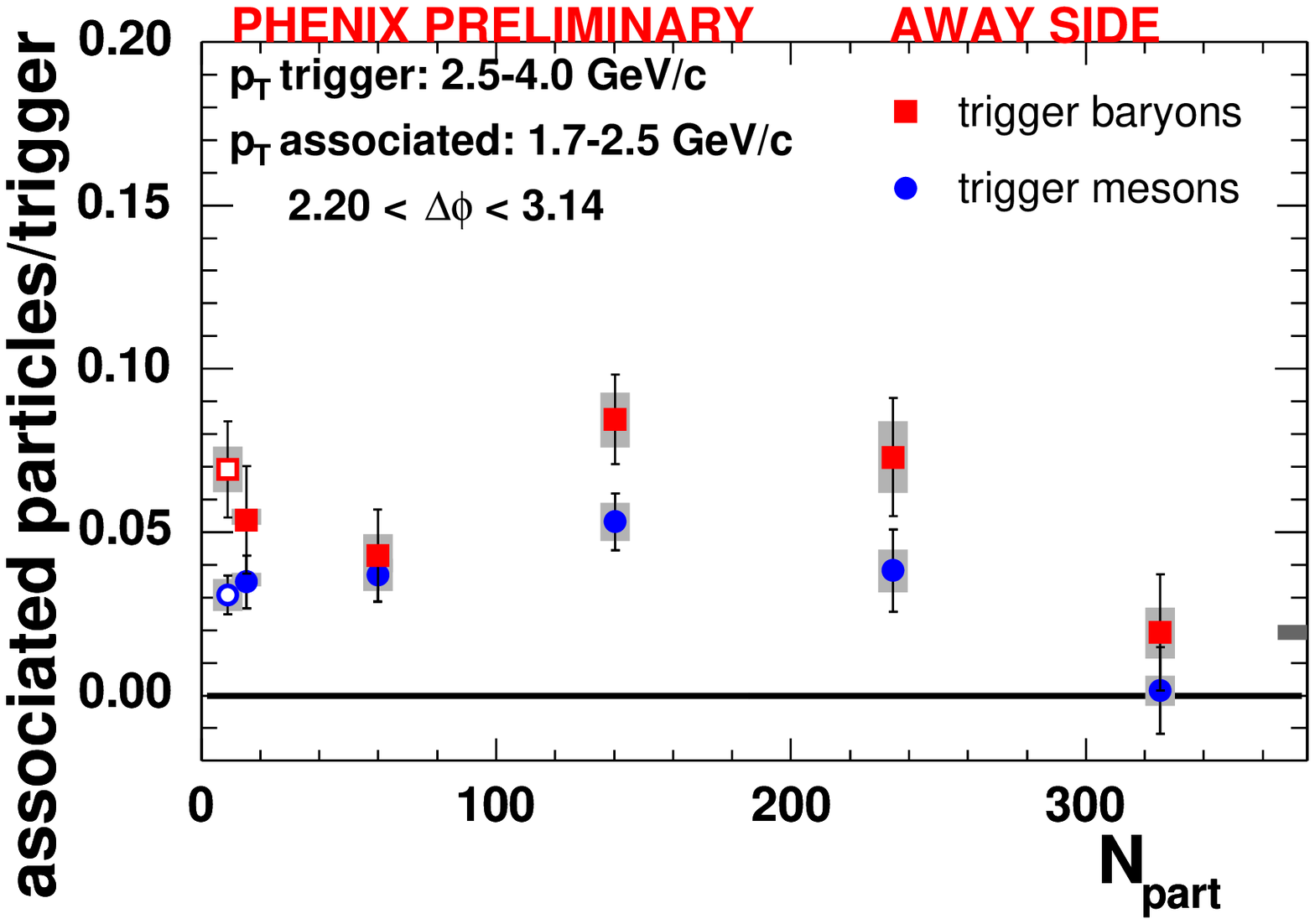}
   \caption{Away side associated particle 
	yield per trigger.  Open points are d+Au values.} 
   \label{idtriga}
   \end{minipage}
\end{figure}

Figure \ref{idtrign} shows the near side associated
particle yield per trigger for both d+Au and
Au+Au as a function of the number of participating nucleons.
There is no decrease in yield going from d+Au to central Au+Au
collisions as would be expected if the triggers (both baryon
and meson) were being increasingly produced by the coalescence of 
thermal quarks.  Within statistical and systematic errors, the centrality
dependence is not strong.

 The fraction of the trigger particle yield arising 
from recombination is taken from \cite{fries}; these triggers should have no 
jet-like partners. The remaining triggers are assigned a partner 
probability equal to that measured in d+Au, shown in the figure by 
the open circles. The combined partner yield per trigger is shown by 
the lines, and clearly under-predicts the data for all 
Au+Au centrality selections. The comparison indicates that the picture is
incomplete, and it has been suggested to include also recombination between 
thermal partons and shower partons created as jets fragments \cite{ko,hwa}.

Figure \ref{idtriga} shows the same quantity for the 
away side jet.  We see a decrease in the yield going
from d+Au to central Au+Au which is in agreement with
previous measurements of away side suppression/broadening
\cite{starb2b,jan}.  

\begin{figure}
 \begin{center}
   \hspace{0.5cm}\includegraphics[width=8.0cm]{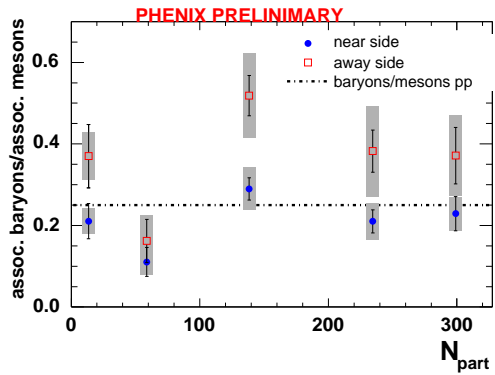}
   \caption{Baryon to meson ratio of jet fragments.  Dashed line
	indicates the baryon to meson ratio from the single
	particle spectra in p+p at the $<p_T>$ of the jet fragments.}
   \label{idassocr}
  \end{center}
\end{figure}

Figure \ref{idassocr} shows the result of the analysis of identified associated particles.
The baryon to meson ratio for associated particles in the same jet as the
trigger particle and the opposing, or away-side jet, are plotted as a function
of the number of participants. The dashed line shows the baryon to meson
ratio of the single particle yields in p+p collisions at the $p_T$ corresponding 
to the mean $p_T$ of the associated particles. The near-side ratio is consistent 
with p+p, but the away-side ratio is higher. Studies are currently underway
to quantify any possible effects of the PHENIX pseudorapidity acceptance upon
this ratio, but the data may be showing a modification of the jet fragmentation by the medium.

In summary, we have shown azimuthal correlations of identified jet fragments.
The identified trigger particle analysis shows
that intermediate $p_T$ meson and baryon production 
is not dominated by the coalescence of thermal partons.  Other models, 
which allow coalescence of thermal partons with those
arising from fragmentation,
may more closely reproduce the data, as a larger
fraction of intermediate $p_{T}$ hadrons would be
predicted to have jet-like partners.  Identifying
the away-side partners offers a way to control the path 
length in the medium and help constrain the models.


\section*{References}

\begin{thebibliography}{99}

\bibitem{phpro} 
PHENIX Collaboration Adler~S~S {\it et al.} 2003
{\sl  Phys.\ Rev.\ Lett.\ }  {\bf 91} 172301

\bibitem{fries} Fries~R~J, M\"{u}ller~B, Nonaka~C and Bass~S~A, 2003 {\sl Phys.\
Rev.\ C} {\bf 68} 044902
\bibitem{hwa} Hwa~R and Yang~C~B 2004 arXiv:nucl-th/0401001
\bibitem{ko} Greco~V, Ko~C~M and L\'{e}vai~P, 2003 {\sl Phys.\ Rev.\ Lett.\ }  {\bf 68} 034904
\bibitem{phenixv2} PHENIX Collaboration Adler~S~S {\it et al.} 2003 {\sl Phys.\ Rev.\ Lett.\ } {\bf 91} 182301
\bibitem{v2react}Bielcikova~J {\it et al.} 2004 {\sl Phys.\ Rev.\ C} { \bf 69} 021901 
\bibitem{starb2b}STAR Collaboration Adler~C {\it et al.} 2003 {\sl Phys.\ Rev.\ Lett.\ }  {\bf 91} 082302
\bibitem{jan} Rak~J for the PHENIX Collaboration, these proceedings.
\bibitem{felix}Matathias~F for the PHENIX Collaboration, these proceedings.
\end{thebibliography}
\end{document}